

\magnification\magstep1
\hsize=37truepc
\hoffset=6pt
\parindent=1.5em
\pageheight{55.5truepc}
\TagsOnRight
\baselineskip=12pt
\NoBlackBoxes
\font\ninepoint=cmr9

\NoBlackBoxes

%
%
%

\def\k{\kappa}
\def\kp{\k_1}
\def\kl{\k_2}

\define\g{{\frak g}}

\define\J{{\Bbb J}}

\define\Ee{{\Bbb E}}
\define\Pe{{\Bbb P}}

\redefine\a{\alpha}
\redefine\be{\beta}

\redefine\t{\theta}

\redefine\C{\text{\ \!C}}

\redefine\S{\text{\ \!S}}
\define\>#1{{\bold#1}}

\define\co{\Delta}
\define\conm#1#2{\left[ #1,#2 \right]}
\define\pois#1#2{\left\{ #1,#2 \right\}}
\def\1{\'{\i}}
\define\q#1{{\left[ #1\right]}_q}
\define\qq#1{{\left\{ #1\right\} }_q}

\def\CGST{1}
\def\FRT{2}
\def\Kup{3}
\def\Mas{4}
\def\BGST{5}
\def\BCG{6}
\def\BHuno{7}
\def\DrYB{8}
\def\Heis{9}

\font\cabeza=cmbx12


\centerline{\cabeza POISSON--LIE CONTRACTIONS AND}
\smallskip
\centerline{\cabeza QUANTUM (1+1) GROUPS}
\bigskip \smallskip

\centerline {A. Ballesteros$^{1),2)}$, F.J.
Herranz$^{1)}$, M.A. del Olmo$^{1)}$ and M. Santander$^{1)}$}
\bigskip
\centerline{\it $^{1)}$Departamento de F\1sica Te\'orica, Universidad
de Valladolid.}
\centerline{\it E-47011 Valladolid, Spain.}
\smallskip
\centerline{\it $^{2)}$Departamento de F\1sica Aplicada III,
Universidad de Valladolid. }
\centerline{\it E.U. Polit\'ecnica, E-09006 Burgos, Spain.}

\medskip
\centerline{ email: fteorica\@cpd.uva.es}

\bigskip

\ninepoint
\noindent ABSTRACT. A Poisson--Hopf algebra of smooth
functions on the (1+1) Cayley--Klein   groups
is constructed by using a classical $r$--matrix which is invariant
under contraction. The quantization of this algebra for the Euclidean,
Galilei and Poincar\'e cases is developed, and their duals are also
computed. Contractions on these quantum groups are studied.

\tenpoint
\medskip
\bigskip\medskip
\medskip

\medskip

The contraction method [\CGST] has already provided a large
number of non--semi\-simple quantum algebras starting from
semisimple ones. However, the contraction of their corresponding $R$
matrices is not possible in general. This fact precludes the use of
the FRT [\FRT] construction to obtain the deformed Hopf
algebra of functions on the group. Recently, the quantization
of (classical) Poisson--Lie (PL) structures has been introduced as an
alternative way to the construction of some non--semisimple quantum
groups [\Kup--\BGST] whose quantum algebra counterparts should be
explicitly derived --when this computation is possible-- as dual Hopf
algebras (this is the case for the Euclidean $E(2)_q$ group
[\Mas,\BCG]). This procedure raises the question about the compatibility
between PL contractions and quantizations, i.e., the commutativity of the
following diagram
$$
\CD Fun(G)
@>{\pois{\,}{\,} \rightarrow \tfrac 1 i\conm{\,}{\,}}>>  Fun(G_q) \\
@V{\varepsilon \rightarrow 0}VV       @V{\varepsilon \rightarrow 0}VV
\\ Fun(G') @>{\pois{\,}{\,} \rightarrow \tfrac 1 i \conm{\,}{\,}}
>>  Fun(G_q') \endCD
$$
where $G'$ is a contraction of $G$ as $\varepsilon\to 0$.

\medskip

A natural framework for studying classical contractions as well as their
quantum versions is the Cayley--Klein (CK) approach [\BHuno]. In (1+1)
dimensions, the (CK) groups $G_{(\k_1,\k_2)}$ are described by means of two
real parameters $(\k_1,\k_2)$. Their corresponding algebras can be
understood as classical deformations of the Galilei one (that has

$(\k_1,\k_2)=(0,0)$). In this context, $\k_1$ (resp. $\k_2$) is a classical
deformation parameter that gives the curvature of the homogeneous space of
points (resp. lines) on which the group acts transitively.

\medskip

In this letter, the Poisson--Hopf algebra $Fun(G_{(\k_1,\k_2)})$ is
constructed. If $\k_1$ vanishes, we obtain the algebra of smooth functions
on the maximal motion groups of the (1+1) dimensional flat spaces of points
(affine (1+1) groups). In this case, the commutativity of the former
diagram relative to the Hopf structure of
$Fun_\varphi(G_{(0,\k_2)})$ and the contraction $\k_2\to 0$ is established.

\medskip

The quantum analogue of the
universal enveloping algebra $U \frak g_{(\k_1,\k_2)}$ has been
given in [\BHuno]:

\smallskip

\noindent {\bf Proposition 1}. {\sl The $\ast$--Hopf algebra
$U_\varphi \frak g_{(\k_1,\k_2)}$ $(\varphi\in\Bbb R)$ is given by
$$
\aligned
\co(P_2)&=1\otimes P_2 + P_2\otimes 1,\\
\co(P_1)&=e^{{\varphi\over 2}
P_2}\otimes P_1 + P_1\otimes e^{-{\varphi\over2}P_2},\\
\co(J_{12})&=e^{{\varphi\over 2} P_2}\otimes J_{12} + J_{12}\otimes
e^{-{\varphi\over 2} P_2};
\endaligned \tag 1
$$
$$
\epsilon(X) =0;\qquad
\gamma(X)=-e^{-{\varphi\over 2}  P_2}\ X\ e^{{\varphi\over 2}  P_2}, \qquad
X\in\{ P_1, P_2, J_{12}\};
\tag 2
$$
$$
\conm{{J}_{12}}{{ P}_1}=i\S_{-\varphi^2}( P_2), \qquad
\conm{{J}_{12}}{{ P}_2}=- i \k_2 { P}_1, \qquad
\conm{{P}_{1}}{{ P}_2}= i \k_1 J_{12}.
\tag 3
$$
The $\ast$--operation is given by
$P_1^\ast=P_1,P_2^\ast=P_2,J_{12}^\ast=J_{12}$.}

\medskip

We recall that the generalized sine and cosine
functions are [\BHuno]
$$
\C_{\k}(x) =
\sum_{l=0}^{\infty} (-\kappa)^l {x^{2l}\over
(2l)!}=
\cases \cos {\sqrt{\k} x} \qquad\quad \text{if }\k>0 \\
1 \qquad \qquad\qquad
\text{if } \k=0 \\
\cosh {\sqrt{-\k} x} \quad\ \  \text{if }\k<0 \\
\endcases
\tag 4.a
$$
$$
\S{_\k}(x) =\sum_{l=0}^{\infty} (-\kappa)^l {x^{2l+1}\over
(2l+1)!}= \cases \frac{1}{\sqrt{\k}} \sin {\sqrt{\k} x}
\qquad\quad\  \text{if }\k>0 \\ x \qquad\qquad\qquad\quad\ \
\text{if } \k=0 \\ \frac{1}{\sqrt{-\k}} \sinh {\sqrt{-\k} x} \quad
\text{if }\k<0 \\ \endcases  .
\tag 4.b
$$
In terms of these functions, a central element for the algebra (3) is
$$
C_\varphi=4 \C_{-\k_1\k_2}({\tfrac \varphi
2})\left[\S_{-\varphi^2}({\tfrac 1 2} P_2) \right]^2 +\tfrac 2
{\varphi} \S_{-\k_1\k_2}({\tfrac \varphi 2}) \left\{ \k_2 P_1^2   +
\k_1 J_{12}^2\right\} .
\tag 5
$$

Provided we
restrict to the affine $\k_1=0$ case, when $\k_2$ is $>0,=0,<0$
we recover the Euclidean $e(2)_q$, Galilei
$g(1+1)_q$  and Poincar\'e $p(1+1)_q$ algebras,
respectively. The generalized (quantum) In\"on\"u--Wigner
contractions are defined by
$$
\alignat5
&\text{ a)}: &\quad\!
\Pe_1&=\varepsilon P_1, &\
\Pe_2&=\varepsilon P_2,&\
\J_{12}&=J_{12},&\  w&={\frac \varphi\varepsilon},\quad
(\varepsilon\rightarrow 0) ;\tag 6\\
&\ \text{b)}:
&\quad\!
\Pe_1&= P_1,&\  \Pe_2&=\varepsilon P_2,&\
\J_{12}&=\varepsilon J_{12},&\  w&={\varphi\over\varepsilon}, \quad
(\varepsilon\rightarrow
0).\tag 7
\endalignat
$$
Transformation (6) (resp. (7)) is completely equivalent to specialize
$\k_1= 0$ (resp. $\k_2=0$) in any expression derived from the CK quantum
(or classical) algebra.

\medskip

\noindent {\bf Lemma 1}. {\sl Under the parametrization
$G_{(\k_1,\k_2)}=e^{a_1 P_1}e^{a_2 P_2}e^{\theta J_{12}}$, a three
dimensional representation of the (1+1) CK groups is given by
$D(G_{(\k_1,\k_2)})=$
\eightpoint
$$
\pmatrix
\C(a_1) \C(a_2) & -\k_1 \S (a_1) \C(\t) -  \k_1\k_2
\C(a_1)
\S(a_2)\S(\t)&\k_1\k_2\S(a_1)\S(\t)- \k_1\k_2
\C(a_1) \S(a_2)\C(\t) \\
\S(a_1) \C(a_2) & \C(a_1) \C(\t) - \k_1\k_2
\S(a_1)
\S(a_2)\S(\t) & - \k_2 \C(a_1)\S(\t)- \k_1\k_2
\S(a_1) \S(a_2)\C(\t) \\
\S(a_2) & \C(a_2) \S(\t)  &
\C(a_2)\C(\t) \endpmatrix.
\tag 8
$$
\tenpoint
For the sake of brevity, we denote $\C(a_1)=\C_{\k_1}(a_1),
\C(a_2)=\C_{\k_1\k_2}(a_2), \C(\t)=\C_{\k_2}(\t)$, and the same convention
is taken for the generalized sine function.

\smallskip

\noindent {\sl Proof.} The fundamental representation $D$
of the classical CK algebra $U\frak g_{(\k_1,\k_2)}$ is
given by
$$
D(P_1)=\pmatrix
0&-\kp&0\\1&0&0\\0&0&0\endpmatrix ,\, D(P_2)=\pmatrix
0&0&-\kp\kl\\0&0&0\\1&0&0\endpmatrix , \, D(J_{12})=\pmatrix
0&0&0\\0&0&-\kl\\0&1&0\endpmatrix.
\tag 9
$$
The expression (8) is readily obtained by computing the
exponentials of the group element and taking into account
definitions (4).

\medskip

Let us now consider the following classical $r$--matrix fulfilling
the modified classical Yang--Baxter equation [\DrYB] for the algebra
$\g_{(\k_1,\k_2)}$
$$
r=\varphi\, J_{12}\wedge P_1.
\tag 10
$$
This element is invariant under both the CK contractions
(6--7).

\medskip

\noindent {\bf Proposition 2}. {\sl The Poisson brackets
$$
\aligned
\pois{a_1}{a_2}&=-\varphi \S_{\k_1}(a_1),\\
\pois{\t}{a_1}&=-\varphi \, {1\over{\C_{\k_1\k_2}(a_2)}} \left(
\C_{\k_2}(\t) - \C_{\k_1}(a_1)\right),\\
\pois{\t}{a_2}&=-\varphi \S_{\k_2}(\t),
\endaligned
\tag 11
$$
give the structure of a Poisson--Hopf
algebra to $Fun(G_{(\k_1,\k_2)})$ .

\smallskip

\noindent Proof:} The Sklyanin bracket induced from a given
$r$--matrix is
$$
\aligned
\pois{f}{g}&=r^{\a\be}\left(X_{\a}^R f X_{\be}^R g -  X_{\a}^L f X_{\be}^L
g \right) \\ &= s\left( X_{J_{12}}^R f X_{P_1}^R g - X_{J_{12}}^L f
X_{P_1}^L g -  X_{P_1}^R f X_{J_{12}}^R g + X_{P_1}^L f X_{J_{12}}^L
g\right). \endaligned
\tag 12
$$

 Left and right invariant vector fields
$X^R,X^L$ are deduced from (8):
$$
\aligned
X_{J_{12}}^L&=\partial_\t,\\
X_{P_1}^L&={1\over{\C_{\k_1\k_2}(a_2)}}\left\{ -\k_1
\S_{\k_1\k_2}(a_2)\C_{\k_2}(\t)  \partial_\t + \C_{\k_2}(\t)
\partial_{a_1} + \C_{\k_1\k_2}(a_2)\S_{\k_2}(\t)
\partial_{a_2}\right\},\\
X_{P_2}^L&={1\over{\C_{\k_1\k_2}(a_2)}}\left\{\k_1\k_2
\S_{\k_1\k_2}(a_2)\S_{\k_2}(\t)  \partial_\t -\k_2 \S_{\k_2}(\t)
\partial_{a_1} + \C_{\k_1\k_2}(a_2)\C_{\k_2}(\t)
\partial_{a_2}\right\}, \endaligned
\tag 13
$$
$$
\aligned
X_{J_{12}}^R &={1\over{\C_{\k_1\k_2}(a_2)}}\left\{ -\k_2
\S_{\k_1\k_2}(a_2)\C_{\k_1}(a_1) \partial_{a_1} + \C_{\k_1}(a_1) \partial_{\t}
+
\C_{\k_1\k_2}(a_2)\S_{\k_1}(a_1) \partial_{a_2}\right\},\\
X_{P_1}^R &=\partial_{a_1},\\
X_{P_2}^R &={1\over{\C_{\k_1\k_2}(a_2)}}\left\{ \k_1\k_2
\S_{\k_1\k_2}(a_2)\S_{\k_1}(a_1)  \partial_{a_1} -\k_1 \S_{\k_1}(a_1)
\partial_{\t} + \C_{\k_1\k_2}(a_2)\C_{\k_1}(a_1)
\partial_{a_2}\right\}. \endaligned
\tag 14
$$
The result (11) is straightforwardly derived from (13--14).

\medskip

By definition, these Poisson brackets are compatible with the group
(co)mul\-tipli\-cation. The contracted PL structures are
immediately obtained by taking a given $\k_i=0$ and come from the
contraction of the group representation (8). The
invariance of $r$ under contractions is essential in order to prevent any
divergency.

\medskip

Hereafter we shall consider the
quantization of the affine $(\k_1=0)$ structures (when both $\k_i$
are non--vanishing, ordering ambiguities appear). In this case, we
assume the standard change of Poisson
brackets (11) into commutation rules
$$
\aligned
[{\hat\theta},{\hat a_1}]&= - i\,\varphi \,(\C_{\k_2}(\hat\theta) - 1),\\
[{\hat\theta},{\hat a_2}]&= - i\,\varphi \,\S_{\k_2}(\hat\theta),\\
[{\hat a_1},{\hat a_2}]&= - i\,\varphi \, \hat a_1.
\endaligned
\tag 15
$$
The contraction ${\k_2}=0$ in (11) gives rise to a formally similar
structure that can be studied in the same way.

\medskip

\noindent {\bf Proposition 3.} {\sl The $\ast$--Hopf algebra
$Fun_\varphi(G_{(0,{\k_2})})$  has multiplication given by
(15), coproduct
$$
\aligned
\co(\C_{\k_2}(\hat\theta))&=\C_{\k_2}(\hat\theta)\otimes\C_{\k_2}(\hat\theta) -
{\k_2} \S_{\k_2}
(\hat\theta)\otimes \S_{\k_2} (\hat\theta),\\
\co(\S_{\k_2}(\hat\theta))&=\S_{\k_2}(\hat\theta)\otimes\C_{\k_2}(\hat\theta) +
\C_{\k_2}
(\hat\theta)\otimes \S_{\k_2} (\hat\theta),\\
\co(\hat a_1)&=\C_{\k_2}(\hat\theta)\otimes \hat a_1 - {\k_2}
\S_{\k_2}(\hat\theta)\otimes \hat a_2 +
\hat a_1\otimes 1,\\
\co(\hat a_2)&=\S_{\k_2}(\hat\theta)\otimes \hat a_1 +
\C_{\k_2}(\hat\theta)\otimes \hat a_2 +
\hat a_2\otimes 1;
\endaligned
\tag 16
$$
counit and antipode
$$
\epsilon(X) =0,\qquad
X\in\{ \hat a_1, \hat a_2, \hat\t\};
\tag 17
$$
$$
\aligned
\gamma(\hat\theta)&=-\hat\theta,\\
\gamma(\hat a_1)&=- \C_{\k_2}(\hat\theta)  \hat a_1 - {\k_2}
\S_{\k_2}(\hat\theta)  \hat a_2, \\
\gamma(\hat a_2)&=\S_{\k_2}(\hat\theta)   \hat a_1 -
\C_{\k_2}(\hat\theta)  \hat a_2, \endaligned
\tag 18
$$
and the $\ast$--operation is
defined by $\hat\t^\ast=\hat\t,\,\hat a_1^\ast=\hat a_1,\,\hat
a_2^\ast=\hat a_2$.}

\medskip

It can be easily shown that $\co (\hat\theta)=1\otimes
\hat\theta + \hat\theta\otimes 1$. Counit $\epsilon$
and antipode $\gamma$ are deduced from the unit matrix and the
inverse element $D(G_{(0,{\k_2})}^{-1})$ (8).

\medskip

The parametrization
we have chosen suggests that the coordinates of the quantum space
$(x_1,x_2)$ should behave like $(\hat a_1,\hat a_2)$. If we
suppose that
$$
\conm{x_1}{x_2}=-i\,\varphi \, x_1,
\tag 19
$$
then the action
$$
\pmatrix 1 \\ x_1' \\ x_2' \endpmatrix= \pmatrix
1 & 0 & 0 \\
\hat a_1 & \C_{{\k_2}}(\hat\t) & - {\k_2} \S_{{\k_2}}(\hat\t) \\
\hat a_2 & \S_{{\k_2}}(\hat\t)  & \C_{\k_2}(\hat\t) \endpmatrix \otimes
\pmatrix 1 \\ x_1 \\ x_2 \endpmatrix
\tag 20
$$
preserves the commutation rules (19) for the $x_i'$ elements.

\medskip

\noindent {\bf Proposition 4.} {\sl The quantum CK(1+1) affine
algebra $U_\varphi \frak g_{(0,{\k_2})}$ is the dual Hopf algebra of
$Fun_\varphi(G_{(0,{\k_2})})$.}

\smallskip

\noindent {\sl Proof:} Let $\Cal A_\varphi$ be the dual Hopf algebra of
$Fun_\varphi(G_{(0,{\k_2})})$. We consider the dual basis
$$
\aligned
\langle E_1, \hat\t^l \hat a_1^m \hat a_2^n \rangle
&=\delta_{l,0}\,\delta_{m,1}\,\delta_{n,0}\\
\langle E_2, \hat\t^l \hat a_1^m \hat a_2^n  \rangle &=
\delta_{l,0}\,\delta_{m,0}\,\delta_{n,1}\\
\langle
E_{12}, \hat\t^l \hat a_1^m \hat a_2^n \rangle &=
\delta_{l,1}\,\delta_{m,0}\,\delta_{n,1} \endaligned \tag 21
$$
and the dual relations
$$
\aligned
\langle A.B, f \rangle &= \langle A\otimes B, \co(f) \rangle,\\
\langle \hat\co(A), f\otimes g \rangle &= \langle A, f\times g
\rangle,\\
\langle A^\ast, f \rangle &=\overline{\langle A, \gamma^{-1}(f^\ast)
\rangle}, \endaligned
\tag 22
$$
where multiplication and comultiplication are denoted $(\times,\co)$ in
$Fun_\varphi(G_{(0,{\k_2})})$ and $(.,\hat\co)$ in $\Cal
A_\varphi$. We obtain the
following commutation rules for the $E$ generators:
$$
\aligned
\conm{{E}_{12}}{{E}_1}&={i\over{2\varphi}}(1-e^{2i\varphi E_2}) -i
 {\k_2}\frac \varphi
2 E_1^2, \quad \\
\conm{{E}_{12}}{{E}_2}&=-{\k_2} {E}_1, \quad \\
\conm{{E}_{1}}{{E}_2}&= 0.
\endaligned
\tag 23
$$
On the other hand, the comultiplication is
$$
\aligned
\hat\co(E_2)&=1\otimes E_2 + E_2\otimes 1,\\
\hat\co(E_1)&=e^{{i\varphi}E_2}\otimes E_1 + E_1\otimes 1,\\
\hat\co(E_{12})&=e^{{i\varphi} E_2}\otimes E_{12} + E_{12}\otimes 1,
\endaligned
\tag 24
$$
and the $\ast$--involution reads
$E_1^\ast=-E_1,\,E_2^\ast=-E_2,\,E_{12}^\ast=-E_{12}$. In view of (23)
and (24) it is immediate to check that the change of basis
$$
P_2=iE_2,\qquad P_1=ie^{-{{i \varphi}\over 2}E_2} E_1, \qquad
J_{12}=ie^{-{{i\varphi}\over 2}E_2} (E_{12} - \tfrac \varphi 4 {\k_2} E_1),
\tag 25
$$
fulfills $P_1^\ast=P_1,P_2^\ast=P_2,J_{12}^\ast=J_{12}$ and leads
to the defining relations of
$U_\varphi \frak g_{(0,{\k_2})}$.

\medskip

In the following we analyze
separately the three quantum groups we have simultaneously
described.

\medskip

\noindent {\it The quantum Euclidean group $Fun_\varphi(G_{(0,1)})$.}
Under the specialization ${\k_2}=1$ and using (4), all the expressions
deduced in a general way are translated into euclidean terms. If we
redefine $t=\hat\theta, n_1=\hat a_1, n_2=-\hat a_2, \varphi =s$, we
recover the $E(2)_q$ quantization obtained in [\BCG]. Recall
that in that work the $R$--matrix method did not give all the
commutation rules: some of them had to be deduced by using
consistency with the group coproduct and with the
$\ast$--involution. This is not the case here, since the $r$--matrix
(10) completely defines the quantization by starting from the
three dimensional real matrix representation (8) (see also [\Mas]).

\medskip

\noindent {\it The quantum Poincar\'e group $Fun_\varphi(G_{(0,-1)})$.}
This case is obtained by taking $\k=-1$ (consequently, rotations are
hyperbolic now). So, we have a quantum algebra of functions generated by
$$
\aligned
[\hat\theta,{\hat a_1}]&= - i\,\varphi \,( \cosh \hat\theta - 1)\\
[\hat\theta,{\hat a_2}]&= - i\,\varphi \,\sinh \hat\theta  \\
[{\hat a_1},{\hat a_2}]&= - i\,\varphi \,\hat a_1,
\endaligned
\tag 26
$$
that has been proven to be the dual of the
quantum (1+1) Poincar\'e algebra.

\medskip

\noindent {\it The quantum Heisenberg group $Fun_\varphi(G_{(0,0)})$.}
In this case relations (15) are
$$
[{\hat\theta},{\hat a_1}]= 0,\qquad
[{\hat\theta},{\hat a_2}]= - i\,\varphi \, \hat\theta, \qquad
[{\hat a_1},{\hat a_2}]= - i\,\varphi \, \hat a_1.
\tag 27
$$
The group element is
$$
T=\pmatrix 1 & 0 & 0 \\
\hat a_1& 1 & 0  \\
\hat a_2 & \hat\theta & 1 \endpmatrix. \tag 28
$$
In fact, relations (27) correspond to the quantum Heisenberg group
obtained from a quasitriangular Hopf algebra [\BGST,\Heis].

\medskip

To close our initial diagram we compute explicitly the
quantum group contraction going from the Poincar\'e (or Euclidean)
case to the Heisenberg one. The pairing between $U_\varphi \frak
g_{(0,{\k_2})}$ and $Fun_\varphi(G_{(0,{\k_2})})$ will be useful at this
point.

\medskip

\noindent {\bf Proposition 5.} {\sl The specialization $\k_2=0$ in
(15--18) is equivalent to the quantum group contraction induced by duality
from the quantum algebra transformation (7).}

\smallskip

\noindent {\sl Proof.} We know that, in  $U_\varphi \frak
g_{(0,{\k_2})}$, the contraction ${\k_2}=0$ is equivalent to a
quantum In\"on\"u--Wigner one given by (7). If we regard (25) as
the defining relations for the $E$ generators, then the latter
inherit a definition for their contraction as follows:
$$
\Ee_1= E_1,\qquad \Ee_2 =\varepsilon E_2,\qquad
\Ee_{12} =\varepsilon E_{12}, \qquad w=\varepsilon^{-1} \varphi,
\qquad (\varepsilon\rightarrow 0).
\tag 29
$$
In turn, duality (21) translates this contraction into the quantum
group generators:
$$
\hat a_1'= \hat a_1,\qquad \hat a_2' =\varepsilon^{-1}
\hat a_2,\qquad  \hat\t' =\varepsilon^{-1} \hat\t,  \qquad
w=\varepsilon^{-1} \varphi, \qquad
(\varepsilon\rightarrow 0).
\tag 30
$$
By applying (30) to relations (26), we
obtain
$$
\aligned
[{\hat\t},{\hat a_1'}]&= -i \, w \,(1  -
\cosh \varepsilon \hat\t'),\\
[{\hat\t},{\hat a_2'}]&= -i\,\varepsilon^{-1}w \sinh
\varepsilon \hat\t',\\
[{ \hat a_1'},{\hat a_2'}]&= - i\,w \,\hat a_1'. \endaligned
\tag 31
$$
In the limit $\varepsilon\rightarrow 0$, (31) gives the Heisenberg
quantum group (27) with deformation parameter $w$. The same is true
for the transformed expressions giving the coproduct, counit and
antipode. Thus, contractions are also described at the quantum
group level by the mere cancellation of a given $\k_i$ parameter
and the full diagram commutes.

\medskip

To end with, some remarks are in order:

\smallskip

\noindent a) Despite of the difficulties that appear in the
quantization of the Hamiltonian structure (11) when both $\k_1,\k_2$
are different from zero, the classical dynamical systems defined by
these brackets are worthwhile to study and will be treated elsewhere.

\smallskip

\noindent b) We emphasize that, throughout this letter, we have been
dealing with classical deformation parameters ($\k_i$) together
with quantum ones ($\varphi$). Both kinds of deformations are
coupled by duality: in $U_\varphi\g_{(0,\k_2)}$, $\k_2$ plays the
role of a structure constant (see (3)) and $\varphi$ is a quantum
parameter that deforms the coproduct (1); on the contrary,
$Fun_\varphi(G_{(0,{\k_2})})$ can be considered as a quantum deformation
of a three dimensional abelian algebra for which the structure
constant is $\varphi$ (compare with (15)) and $\k_2$
arises as the quantum parameter deforming the Hopf algebra structure
(16). In this sense, the CK scheme can be helpful to make clear this
kind of correlation between classical and quantum deformations,
since it enables both of them to be simultaneoulsly described for a
rather representative family of algebras.

\bigskip
\bigskip

\noindent {\bf Acknowledgments}
\medskip

We acknowledge Prof. P.P. Kulish for helpful discussions. This work
has been partially supported by a DGICYT project (PB91--0196) from
the Ministerio de Educaci\'on y Ciencia de Espa\~na.

\bigskip

\noindent\bf References
\rm
\eightpoint

\smallskip

\ref     
\no{[\CGST]}
\by      Celeghini E.,  Giachetti R.,  Sorace E., and  Tarlini M.
\paper   Contractions of quantum groups
\jour    Lecture Notes in Mathematics n. 1510, 221,
         Springer-Verlag, Berl\1n (1992)
\endref

\ref     
\no[{\FRT}]
\by      N. Yu. Reshetikhin, L.A. Takhtadzhyan and L.D. Faddeev
\jour    Leningrad Math. J.
\vol     1
\yr      1990
\pages   193
\endref

\ref     
\no[{\Kup}]
\by      B.A. Kuperschmidt
\jour    J. Phys. A: Math. Gen
\vol     26
\pages   L929
\yr      1993
\endref

\ref     
\no[{\Mas}]
\by       P. Maslanka
\paper    The $E_q(2)$ group via direct quantization of
Lie--Poisson structure and its Lie algebra
\jour     to appear in J. Math. Phys
\endref

\ref     
\no[{\BGST}]
\by      F. Bonechi, R. Giachetti, E. Sorace, and M. Tarlini
\paper   Deformation Quantization of the Heisenberg Group
\jour    DFF preprint
\yr      1993
\endref

\ref     
\no[{\BCG}]
\by       A. Ballesteros, E. Celeghini, R. Giachetti, E. Sorace and
M. Tarlini
\jour     J. Phys. A: Math. Gen.
\vol      26
\yr      1993
\pages   7495
\endref

\ref        
\no[{\BHuno}]
\by         A. Ballesteros, F.J. Herranz, M.A. del Olmo and M. Santander
\jour       J. Phys. A: Math. Gen.
\vol        26
\yr         1993
\pages      5801
\endref

\ref        
\no[{\DrYB}]
\by         V.G. Drinfel'd
\jour       Soviet Math. Dokl.
\vol        27
\yr         1983
\pages      68
\endref

\ref        
\no[{\Heis}]
\by         A. Ballesteros, E. Celeghini, F.J. Herranz, M.A. del Olmo and
M. Santander
\paper      A universal non--quasitriangular quantization of the
Heisenberg group
\jour       UVA preprint
\yr         1994
\endref

\end